\documentclass[11pt,a4paper]{article}
\usepackage{amsmath,amssymb,amsfonts,slashed}
\usepackage[text={176mm,240mm},centering]{geometry}
\usepackage{float}
\usepackage[figuresright]{rotating}
\usepackage{multirow}
\usepackage{color}
\usepackage[colorlinks=true,linkcolor=red,breaklinks=true,urlcolor=blue,citecolor=blue]{hyperref}
\usepackage{cite}

\newcommand \bra [1] {\langle {#1}\vert}
\newcommand \ket [1] {\vert {#1}\rangle}

\def\be{\begin{equation}}
\def\ee{\end{equation}}
\def\ba{\begin{eqnarray}}
\def\ea{\end{eqnarray}}
\def\3P0{{}^3P_0}
\begin{document}
\title{How Large is the Contribution of Excited Mesons in Coupled-Channel Effects?}

\date{\today}

\author{
     Yu Lu$^{1,3}$
                         \footnote{Email address:\texttt{luyu211@ihep.ac.cn} }~ ,
     Muhammad Naeem Anwar $^{2,3}$
                        \footnote{Email address:\texttt{naeem@itp.ac.cn} }~ ,
      Bing-Song Zou$^{2,3}$
                        \footnote{Email address:\texttt{zoubs@itp.ac.cn}}
       \\[2mm]
      {\it\small$^1$Institute of High Energy Physics, Chinese Academy of Sciences, Beijing 100049, China,}\\
      {\it\small$^2$CAS Key Laboratory of Theoretical Physics, Institute of Theoretical Physics,}\\
      {\it\small  Chinese Academy of Sciences, Beijing 100190, China}\\
      {\it\small$^3$University of Chinese Academy of Sciences, Beijing 100049, China}\\
}

\maketitle

\begin{abstract}
We study the excited $B$ mesons' contributions to the coupled-channel effects under the framework of the ${}^3P_0$ model for the bottomonium.
Contrary to what has been widely accepted,
the contributions of $P$ wave $B$ mesons are generally the largest,
and to some extent, this result is independent of the potential parameters.
We also push the calculation beyond $B(1P)$ and carefully analyze the contributions of $B(2S)$.
A form factor is a key ingredient to suppress the contributions of $B(2S)$ for low lying bottomonia.
However, this suppression mechanism is not efficient for highly excited bottomonia,
such as $\Upsilon(5S)$ and $\Upsilon(6S)$.
We give explanations why this difficulty happens to the ${}^3P_0$ model and
suggest analyzing the flux-tube breaking model for the full calculation of coupled-channel effects.
\end{abstract}

\thispagestyle{empty}

\section{Introduction}

Heavy quarkonium is a multiscale system covering all regimes of quantum chromodynamics (QCD),
which make it an ideal place to study strong interactions~\cite{Brambilla:2010cs}.
Despite the success of QCD in the high energy region, due to asymptotic freedom,
the nonperturbative effect dominates at low energies and brings problems to perturbative calculation.
One tool to study this non-perturbative effect is lattice QCD.
However, due to its huge calculation work, it is still unable to calculate all the physical quantities with the current computation power.
Another important approach is to develop various phenomenological models.
Among these phenomenological models, the quark model is a prominent one.
Under the quark model framework,
various types of interactions have been suggested by various groups,
and they have achieved many impressive successes (see e.g. Refs.~\cite{Martin:1980rm,Bertlmann:1979zs,Eichten:1978tg,Buchmuller:1980su,Godfrey:1985xj}).
However, these potential models cannot be the whole story.
One important missing ingredient is the mechanism to generate quark-antiquark pairs
which enlarge the Fock space of the initial state,
i.e., the initial state contains multiquark components.
These multiquark components will change the Hamiltonian of the potential model,
causing mass shift and mixing between states with the same quantum numbers
or directly contributing to an open channel strong decay if the initial state is above the corresponding threshold.
These consequences can be summarized as unquenched effects or coupled-channel effects.

Through various approaches,
such as the $\3P0$ model~\cite{Micu:1968mk,LeYaouanc:1972ae,LeYaouanc:1973xz,Blundell:1996as,Barnes:2007xu},
flux-tube breaking model~\cite{Carlson:1982xi,Carlson:1983rw,Kokoski1987,Geiger:1989yc,Close:1994hc},
microscopic decay models~\cite{Eichten:1978tg,Eichten:1979ms,Ackleh:1996yt,Barnes:2005pb},
$S$ matrix analysis~\cite{Baru:2010ww,Hanhart:2011jz,Guo:2016bjq,Hammer:2016prh},
coupled-channel effects are extensively studied in many literatures (e.g., Refs.~\cite{Heikkila:1983wd,Ono:1983rd,Tornqvist:1984fy,Ono:1985eu,Ono:1985jt,Danilkin:2009hr})

Despite these pioneering works,
we find that at least two factors have the potential to jeopardize the calculations of the coupled-channel effects.
One is the widely used simple harmonic oscillator (SHO) wave function
which approximates the realistic wave function in the wave function overlap integration.
As already pointed out in our previous work \cite{Lu:2016mbb},
the SHO approximation is not good, especially for states near thresholds,
and it is essential to treat the wave functions precisely.

Another is the assumption that the contribution of the excited meson loops is negligible.
Take the bottomonium, which is the system we study in this paper, as an example.
Two reasons may explain why this approximation is widely used.
One is that the threshold corresponding to the excited $B$ mesons is higher than the ground state $B\bar{B}$ threshold;
thus the coupled-channel effects is expected to be small.
Another is that the calculations of the excited $B$ mesons' contributions are more complicated,
an effective method to do the calculation is still not widely known.

As explained in\cite{Lu:2016mbb},
the Gaussian expansion method plus the techniques of transformation between the Cartesian and spherical basis can,
in principle, do the sophisticated calculations.
Another observation is that the first excited $P$ wave $B$ meson
is only around $450$MeV heavier than the ground state $B$ meson.
This mass difference is less than one tenth of the $B$ mesons' mass.
Given the fact that the quantum numbers of $B_1$ are different from the ground state $B$ mesons,
and the coupled-channel effects do rely on the quantum numbers,
the suppression purely originates from the larger mass may be not as large as what has been taken for granted.

The sum of all the intermediate meson loops is more than just a calculation challenge,
it may also lead to profound physics.
In the light sector,
a series of works by Geiger \emph{et al.}~\cite{Geiger:1991qe,Geiger:1991ab,Geiger:1992va}
show us that even though different intermediate meson loops contribute to the breaking of Okubo-Zweig-Iizuka (OZI) rule,
under some simplifications (such as neglecting the mass difference in the denominator),
they contribute destructively, leaving us a perfect OZI rule.
If one only sums over some of the meson loops, one may leave with a wrong conclusion.

One should not confuse their calculation with what we are going to do in this work.
They studied the flavor-changing process,
such as $u\bar{u} \to \text{virtual meson pairs} \to d\bar{d}$.
However, we mainly focus on the mass shift and their cancellation does not happen to our case.
For states below the threshold, the intermediate loops always contribute a negative mass shift.
Even though for the above threshold case,
where the mass shift may add destructively,
the mass difference does matter in the real calculation.

In Ref.~\cite{Geiger:1989yc},
under the flux-tube breaking model,
Geiger and Isgur also showed that if one adds up all the intermediate states,
the mass shift caused by coupled-channel loops does not converge when two assumptions are adopted;
i) the meson wave function is the SHO wave function,
and ii) the string length between the generated quark pairs is zero.
In the $\3P0$ model, this conclusion needs to be checked or recalculated,
because the SHO approximation is far from true, which was already stressed before
and the vertex of the quark generation is different from the flux-tube breaking model.

There are also some studies of the excited meson loops for heavy quarkonium.
In the charmonium sector,
a lattice calculation in Ref.~\cite{Bali:2011rd} shows that
$\eta_c$ and $J/\psi$ has small but non-negligible components of $D_1\bar{D}^*$.
In bottomonium sector,
loops involving a $B_1$ meson are proven to be critical to explain
the large breaking of the heavy quark spin symmetry of $\Upsilon(10860)$
\cite{Guo:2014qra}.
However, these studies focus on some specific states,
and systematic studies of the excited meson loops is still missing.

To summarize,
it remains to be answered or clarified whether the ground state approximation is good or not,
and the general properties of the excited meson loops still need to be systematically exploited.
In this work,
we try to answer these questions under the framework of the $\3P0$ model.

This paper is organized as follows.
In Sec.~\ref{modelSec},
we briefly describe the ingredients of the coupled-channel effects and the calculation techniques,
including the $\3P0$ model, the Cornell potential model,
the Gaussian expansion method, and the transformations between the spherical and Cartesian basis.
Sec.~\ref{resultSec} is devoted to the results of the higher excited $B$ mesons up to $B(2S)$,
and we explain the necessity of the form factor and the limitation of the $\3P0$ model.
Finally, we give a short summary in Sec.~\ref{summary}.

\section{Theoretical Framework}\label{modelSec}
The calculation methods and tools are described in details in Ref.~\cite{Lu:2016mbb};
thus, we only sketch the key steps in this section.

\begin{figure}[h]
  \centering
  \includegraphics[width=0.6\textwidth]{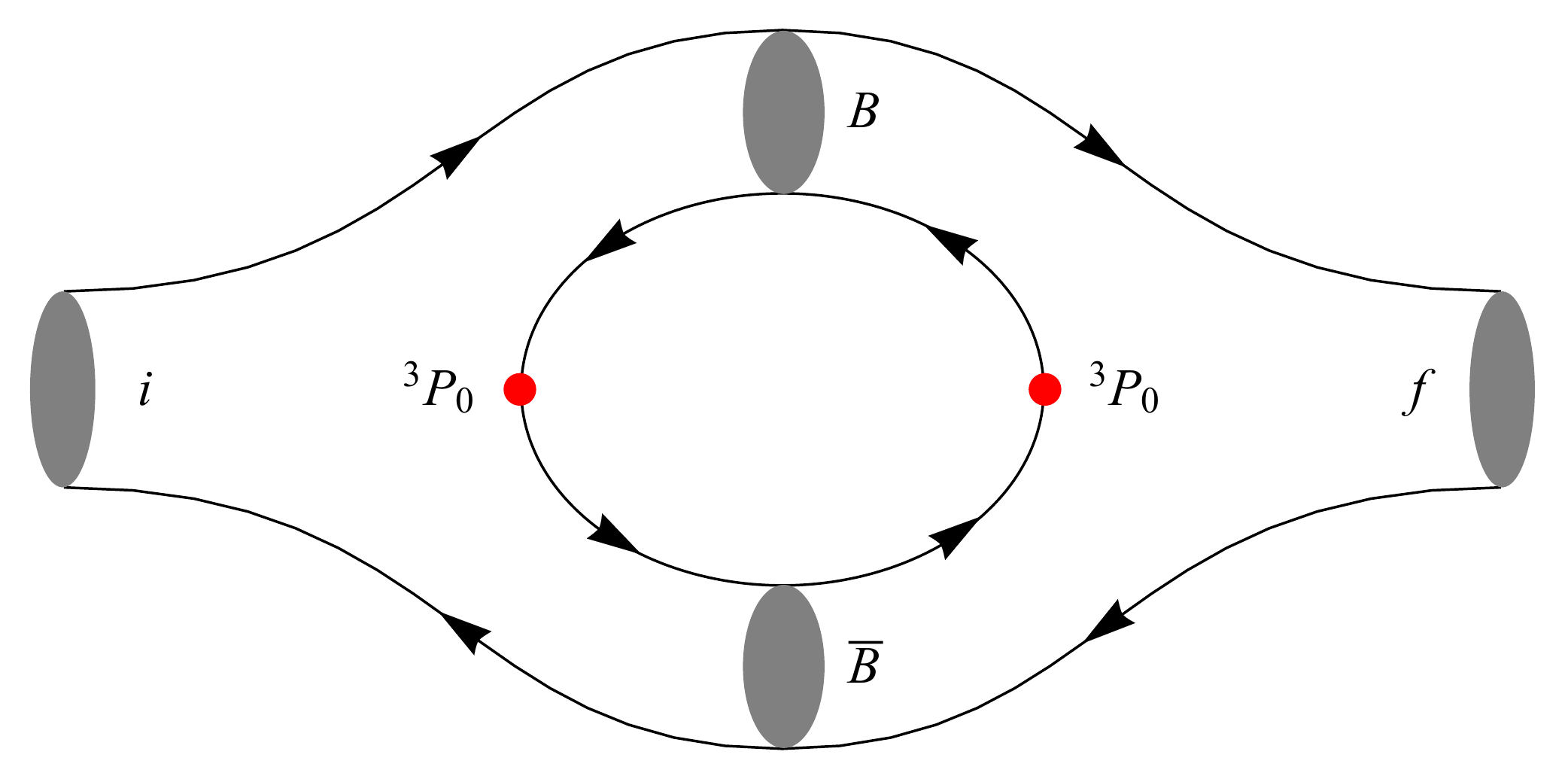}\\
  \caption{Sketch of coupled-channel effects in the $\3P0$ model for the bottomonium.
   $i$ and $f$, respectively, denote the initial and final states with the same $J^{PC}$ and
   $B\bar{B}$ stands for all possible $B$ meson pairs, including the excited $B$ mesons.
  }
 \label{CCDiagram}
\end{figure}

The coupled-channel effects in the $\3P0$ model
\cite{Micu:1968mk,LeYaouanc:1972ae,LeYaouanc:1973xz} can be described by Fig.~\ref{CCDiagram}.
In this model, the generated quark pairs have a vacuum quantum number $J^{PC}=0^{++}$.
In the notation of ${}^{2S+1}L_J$, it is $\3P0$ that explains the model's name.

The Hamiltonian to generate the quark-antiquark pairs is expressed as
\be
H_I=2 m_q \gamma \int d^3x \bar{\psi}_q \psi_q,
\ee
where $m_q$ is the produced quark mass and$\gamma$ is the dimensionless coupling constant.
Since the probability to generate heavier quarks is suppressed,
we use the effective strength $\gamma_s=\frac{m_q}{m_s}\gamma$ in the following calculation,
where $m_q=m_u=m_d$ is the constituent quark mass of the up (or down) quark and $m_s$  is the strange quark mass.

As sketched in Fig.~\ref{CCDiagram}, the experimentally observed state
should be a mixture of a pure quarkonium state $\ket{\psi_0}$ and a $B$ meson continuum state $\ket{BC;p}$.
The experimentally observed state $\ket{A} $ should be expressed as
\be
\ket {A}=c_0 \ket{\psi_0} +\sum_{BC} \int d^3p\, c_{BC}(p) \ket{BC;p},
\ee
which is the eigenstate of the full Hamiltonian $H$ defined in Eq.~\ref{fullH}.
$c_0$ and $c_{BC}$ stand for the normalization constants of the bare state and $B$ meson continuum, respectively.

In order to work out $\ket {A}$, as a first step,
one has to solve the wave function $|\psi_0\rangle$ for the heavy quarkonium.
In this work, it is obtained by solving the Schr\"{o}dinger equation
\be
H_0\ket{\psi_0}=\left(2m_b+\frac{p^2}{m_b}+V(r)+V_s(r)\right)\ket{\psi_0}=M_0 \ket{\psi_0}
\ee
where $m_b$ and $M_0$ represent the mass of $b$ quark and the bare mass of the bottomonium, respectively.
In the above equation, $V(r)$ is the well-known Cornell potential~\cite{Eichten:1978tg,Eichten:1979ms}
\be
V(r)=-\frac{4}{3} \frac{\alpha}{r}+\lambda r+c,
\label{Potential}
\ee
where $\alpha, \lambda$ and $c$ stand for the strength of color Coulomb potential,
the strength of linear confinement and mass renormalization, respectively.
$V_s(r)$ stands for the spin dependent interactions which restores the hyperfine or fine structures of the bottomonium,

\be
V_{s}(r)=\left(\frac{2\alpha}{m^2_b r^3}-\frac{\lambda}{2m^2_b r}\right)\vec{L}\cdot \vec{S}
+\frac{32\pi \alpha}{9m_b^2}\tilde{\delta}(r) \vec{S}_b\cdot \vec{S}_{\bar{b}}
+\frac{4\alpha}{m^2_b r^3}\left(\frac{\vec{S}_b\cdot \vec{S}_{\bar{b}}}{3}
+\frac{(\vec{S}_b\cdot\vec{r}) (\vec{S}_{\bar{b}}\cdot \vec{r})}{r^2}\right),
\label{finestructure}
\ee
where $\vec{L}$ denotes relative orbital angular momentum,
$\vec{S}=\vec{S}_b+\vec{S}_{\bar{b}}$  is the total spin of the $b$ quark pairs
and $m_b$ is the $b$ quark mass.
$\tilde{\delta}(r)$ is the smeared delta function and is written as
$\tilde{\delta}(r)=(\sigma/\sqrt{\pi})^3 e^{-\sigma^2 r^2}$~\cite{Barnes:2005pb,Li:2009ad}.
We treat the spin dependent term as a perturbation,
and the spatial wave functions are obtained by solving Schr\"{o}dinger  equation numerically using Numerov's method~\cite{Numerov:1927}.

Combine the Cornell potential and the dynamics of quark pair generation,
we get the full Hamiltonian,
\be
  H=H_0+E_{BC}+H_I,\label{fullH}
\ee
where $E_{BC}=\sqrt{m_B^2+p^2}+\sqrt{m_C^2+p^2}.$
The $H_I$ term in Eq.~\ref{fullH} is mainly responsible for the mass shift.
As the name tells us, it is naturally defined as
\be
\Delta M \equiv M-M_0 \label{intEqn}
\ee
and it can be obtained by solving the integral equation
\be
\Delta M=\sum_{BC}\int d^3p\, \frac{\vert \bra {BC;p} H_I \ket{\psi_0} \vert ^2}{M-E_{BC}-i\epsilon}. \label{mShift}
\ee
Note that the $i\epsilon$ term is added to handle the situation when $m_A> m_B+m_C$.
In this case, $\Delta M$ will pick up an imaginary part
\be
\mathrm{Im} (\Delta M)=\sum_{BC} \pi P_B \frac{E_B E_C}{m_A} \vert \bra {BC;P_B} H_I \ket{\psi_0} \vert ^2,\label{decay}
\ee
which is equal to one half of the the decay width.
$P_B$ and $ E_B$ denote the momentum and energy of $B$ meson, respectively.
The wave function overlap integration lies in the term
\be
\bra{BC;P_B} H_I \ket{\psi_0}=
\sum_{\text{polarization}}\int d^3k
\phi_0(\vec{k}+\vec{P}_B) \phi_B^*(\vec{k}+x \vec{P}_B)\phi_C^*(\vec{k}+x \vec{P}_B)
|\vec{k}| Y_1^m(\theta_{\vec{k}},\phi_{\vec{k}}),\label{overlap}
\ee
where $x=m_q/(m_b+m_q)$ and $m_b$ and $m_q$ denote the $b$ quark and the light quark mass, respectively.

Once $M$ is solved, the coefficient of different components can be worked out either.
For states below the threshold, the probability of the $b\bar{b}$ component is expressed as
\be
P_{b\bar{b}}:=\vert c_0 \vert ^2=\left(1+\sum_{BC LS}\int_0^\infty dp \frac{p^2\vert \mathcal{M}^{LS}\vert^2}{(M-E_{BC})^2}\right)^{-1}, \label{1prob}
\ee
where $\vert \mathcal{M}^{LS}\vert^2$ is represented as
\be
\vert \mathcal{M}^{LS}\vert^2=\int d\Omega_B \, \vert \bra {BC;P_B} H_I \ket{\psi_0} \vert ^2.
\ee

The main calculation work lies in the Eq.~\ref{overlap}.
In order to evaluate it precisely,
we use the Gaussian expansion method (GEM)\cite{Hiyama:2003cu} and
the transformation between the spherical and Cartesian basis to make the GEM automatic applicable to the excited $B$ meson.

\section{Results and Discussions}\label{resultSec}
\subsection{$B_1$'s Contributions with Traditional $\3P0$ Model}
In this work, we do not intend to reproduce the spectrum and decay widths of the bottomonium family,
but to study the general contributions of the excited B meson to coupled-channel effects.

There are two simplifications in our calculation.
Firstly, since the constituent quark mass of $u, d$ quarks are set to be the same to get the wave functions,
we further make the approximation $m(B^0)\approx m(B^\pm)$.
Notwithstanding this simplification, there are still 42 channels to be calculated,
i.e., $B_{(s)}^{(*)} \bar{B}_{(s)}^{(*)}, B_{(s)}^{(*)} \bar{B}_{s}(1P)$ and $\bar{B}_{(s)}(1P) \bar{B}_{(s)}(1P)$.
In principle, one can still treat their mass precisely,
however, based on our experiences,
this simplification is quite precise, and it saves a lot of the calculation work.

Second simplification comes from $B_1$ meson multiplets.
From the perspective of the quark model, there are four $1P$ wave $B$ mesons.
Nevertheless, the predicted $B_0(1P)$ and one of the $B_1(1P)$  are still not experimentally observed.
As far as the coupled-channel is concerned,
the missing of these two states is not a real problem.
Because the mass of $b$ quark is very large,
it is safe to assume that the heavy quark spin symmetry does not break.
Under this limit,
it is reasonable to set the mass of $B_0({}^3P_0)$  and missing partner of $B_1(5721)$  to be the same as $B_1(5721)$.
There is a mixing between $B({}^3P_1)$ and $B({}^1P_1)$ to form the experimentally observed $B_1(5721)$\cite{Olive:2016xmw}.
However, since we sum up all the possible combinations of $B$ mesons,
mixing between the two $1^+ B$ mesons can also be neglected.
As far the mass of the intermediate states are well measured,
we use the value from the Particle Data Group~\cite{Olive:2016xmw}.

\begin{figure}[H]
  \centering
  \includegraphics[width=\textwidth]{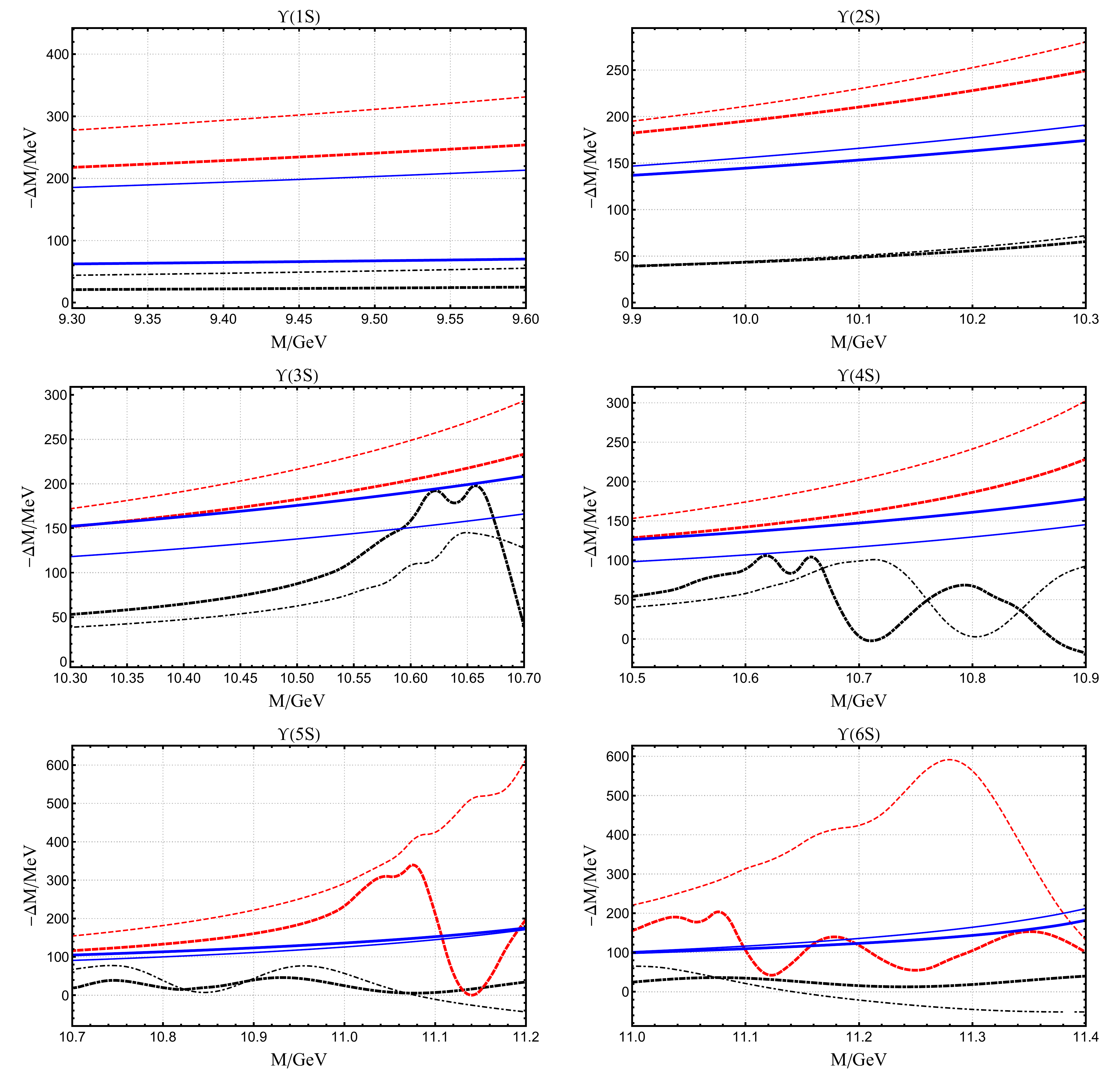}\\
  \caption{$-\Delta M$ of different coupled-channels for $\Upsilon(nS)$.
  The results with parameters in Tab.~\ref{paraTab} and the results recalculated with parameters of Ref.~\cite{Liu:2011yp}
  are represented by thick and thin curves, respectively.
  $\Delta M(1S,1S), \Delta M(1S,1P)$, and $ \Delta M(1P,1P)$ are represented by
  black dot-dashed, red dashed and blue solid curves, respectively.
  }
 \label{ShiftCompare}
\end{figure}

The coupled-channels are much more involved than the case,
which only includes the ground state $B$ mesons.
For illustration purposes,
we sum up the contributions of the states in the same multiplet and $u, d, s$ flavors and their charge conjugate partners,
classifying the channels into three groups (see Fig~\ref{ShiftCompare}),
e.g., $B(1S)$ stands for ground state $B,B^*,B_s$, and $B_s^*$ mesons.
Also, in the rest of this paper,
we define $\Delta M(i,j)$ to represent the sum of the mass shift due to all the possible combinations of $i$ and $j$ wave $B$ mesons.

The parameters are given in Table~\ref{paraTab},
which are the same as our previous work~\cite{Lu:2016mbb}.
\begin{table}[H]
  \renewcommand\arraystretch{1.1}
  \centering
\begin{tabular}{cccccccccc}
 \hline\hline
$\alpha=0.34$   & $\lambda=0.22\text{GeV}^2$    &$c=0.435\text{GeV} $\\
$m_b=4.5\text{GeV}$ & $m_u=m_d=0.33\text{GeV}$& $m_s=0.5\text{GeV}$\\
$\sigma=3.838\text{GeV}$& $\gamma=0.205$ \\
 \hline \hline
\end{tabular}
\caption{The parameters used in our calculation.}
\label{paraTab}
\end{table}

The mass shift $\Delta M$ of both $B(1S)$ and $B(1P)$ mesons are depicted in Fig.~\ref{ShiftCompare}.
In the rest of this section,
we are going to discuss some interesting structures of the plots.

When the states are far below the threshold,
the mass shift is always negative [as indicated in Eq.~\ref{mShift}],
and closer to the threshold means an increase of $|\Delta M|$.
When the mass goes higher,
the magnitude of $\Delta M$ is no longer monotonically increasing but oscillating,
which reflects the node structure of the wave functions.
This phenomenon happens to both cases whether $B_1$ mesons' contributions are considered or not.

A specific example would help to clarify this point.
For $\Upsilon(10860)$, which we treat it as $\Upsilon(5S)$,
 $\Delta M(1S,1P)$ is generally the largest
but there is a big dip around $11.14$ GeV with parameters in Tab.~\ref{paraTab};
however, in Ref.\cite{Liu:2011yp}'s parameters, this channel's contribution always grows.
In these cases, it is difficult to make a solid conclusion about the spectrum behavior,
i.e., the spectrum is sensitive to the potential parameters.
This conclusion also agrees with Ref.~\cite{Tornqvist:1984fx}.

The most notable impact of $B_1$ family is the unexpected large contributions to the mass shift,
and generally, $\Delta M(1S,1P)$ is the largest.
Even for $\Upsilon(1S)$, which is far below the threshold,
$\Delta M(1S,1P)+\Delta M(1P,1P)$ is 14 times larger than $\Delta M(1S,1S)$.

Compared with the case that only considers the ground states $B$ mesons contributions,
threshold effect is more clearly reflected when the excited $B$ mesons is included,
because open channels due to different multiplets are well separated.

\begin{figure}[H]
  \centering
  \includegraphics[width=0.9\textwidth]{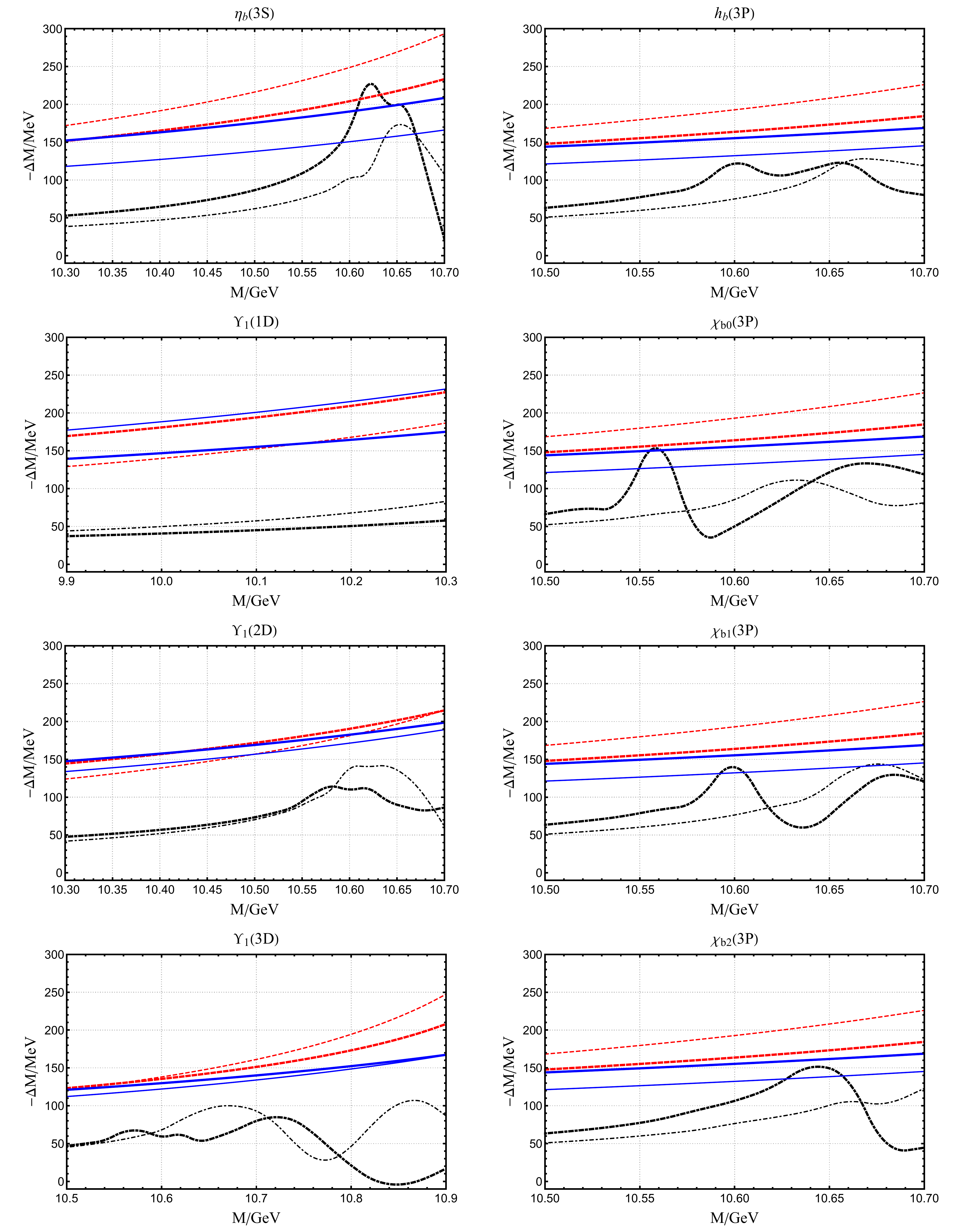}\\
  \caption{$-\Delta M$ of different coupled-channels for some selected representatives of the
  ${}^1S_0,{}^1P_1,{}^3P_J,{}^3D_1$ families.
  The results with parameters in Tab.~\ref{paraTab},
  and results recalculated with parameters of Ref.~\cite{Liu:2011yp}
  are represented by thick and thin curves, respectively.
  $\Delta M(1S,1S), \Delta M(1S,1P)$, and $ \Delta M(1P,1P)$ are represented by
  black dot-dashed, red dashed, and blue solid curves, respectively.
  Here we omit the results of $\eta_b(nS), h_b(nP)$, and $\chi_{bj}(nP)$ when $n\le 2$,
  because these results are quite similar to those of $\Upsilon(1S)$ or $\Upsilon(2S)$.
  }
 \label{ShiftCompare2}
\end{figure}

The first open bottom threshold is $2m_B\approx10.56$GeV.
When the mass approaches to this value,
the denominator $M-E_{BC}$ approaches to 0 in Eq.~\ref{mShift};
thus, there is an enhancement of $\Delta M$ from the $B(1S)\bar{B}(1S)$ channel.
From the $\Upsilon(3S)$ case in Fig.~\ref{ShiftCompare},
one can clearly see the sharp increase of $\Delta M(1S,1S)$ between $10.55$ and $10.6$GeV,
while the slop of the other channels does not change much because their threshold is 450MeV larger.

The threshold effect is broken in some degree by the nodes of the wave functions.
The peaks and valleys of the wave function are more likely to cancel with each other for higher excited states,
leaving a relatively small slope of $\Delta M$.
One can compare the result of $\Upsilon(4S)$ with $\Upsilon(3S)$ verify this conclusion.

We need to stress that ${}^3S_1$ is not the only family who couples strongly with $B\bar{B}(1P)$ loop.
As shown in Fig~\ref{ShiftCompare2},
all the families of ${}^1S_0,{}^1P_1,{}^3P_J,{}^3D_1$ share these general properties.

An direct consequence of the large $\Delta M(1S,1P)$ is that
the parameters only considering ground state $B$ mesons
to reproduce the experimental data are somewhat incomplete or even misleading.

This conclusion is independent of the parameters to some degree,
since the results are based on two different sets of parameters (Refs.~\cite{Lu:2016mbb} and ~\cite{Liu:2011yp}),
both give large $\Delta M(1S,1P)$.

It seems feasible that to fit the spectrum,
one may tune the parameters,
such as the mass renormalization term $c$ in Eq.~\ref{Potential},
to absorb this unexpected large mass shift,
or take a further step,
weaken the confine potential to reflect the fact that
closer to the threshold generally means more suppression of the mass.

This could be a possible way out,
if someone only focuses on the spectrum.
However, as pointed out in our previous work~\cite{Lu:2016mbb},
the mass shift only reveals one aspect of the coupled-channel effects.

To be specific, the big contribution of $B(1P)$ not only brings us the large mass shift $\Delta M$,
but also the large fraction of meson pairs (or equivalently, small $P_{b\bar{b}}$).
This renormalization of the wave function cannot be accommodated in the framework of the potential model.

\subsection{Beyond $B_1$}\label{B1Section}

The large contribution of $B_1$ mesons reminds us that
it is essential to do the \emph{ab initio} calculations of coupled-channel effects.
However, before really carrying out these calculations,
one may naturally ask this question,
``how large are the contributions of the states beyond $B_1$ or
how to evaluate all the intermediate meson loops?"

It is very difficult to offer a complete answer to this question.
Nevertheless, it turns out that we can still estimate the contributions up to $B(2S)$.
Furthermore, by analyzing the mass shift behavior carefully,
we can draw some model independent conclusions.

First of all, we are trying to precisely evaluate the loops.
Of course, since the $B(2S)$ mass and its wave functions are not known,
the calculation has to be model dependent, which is inevitable.

So far, the $B(2S)$ meson has not been experimentally well determined and its mass is model dependent.
For our purposes,
it is sufficient to take an average of several theoretical estimations in
Refs.~\cite{Zeng:1994vj,DiPierro:2001dwf,Ebert:2009ua,Lahde:1999ih,Lu:2016bbk},
and for the consistency treatment of the wave functions,
the $B(2S)$ wave functions are still deduced from our parameters.

We define the ratio
\be
R:=\frac{\Delta M(2S,2S)+\Delta M(2S,1P)+\Delta M(2S,1S)}{\Delta M(1S,1S)+\Delta M(1S,1P)+\Delta M(1P,1P)}
\label{Rdefine}
\ee
to tell how large are the contributions of $B(2S)$.

As will be explained below, in the calculation when $B(2S)$ are involved,
we adopt the Gaussian form factor which reveals the size of the generated quark pairs
(see, e.g., Ref.~\cite{SilvestreBrac:1991pw,Roberts:1997kq,Ferretti:2013vua}).
This form factor modifies $\bra{BC;P_B} H_I \ket{\psi_0}$ to be
\be
\bra{BC;P_B} H_I \ket{\psi_0}=
\sum_{\text{polarization}}\int d^3k e^{-2 r^2 k^2/3}
\phi_0(\vec{k}+\vec{P}_B) \phi_B^*(\vec{k}+x \vec{P}_B)\phi_C^*(\vec{k}+x \vec{P}_B)
|\vec{k}| Y_1^m(\theta_{\vec{k}},\phi_{\vec{k}}),\label{overlapFF}
\ee

In this work, we fit $r$ to be $0.408$ fm, which minimizes $R$ of $\Upsilon(1S)$.
This value slightly larger than the value $r=0.335$ fm used in Ref.~\cite{Ferretti:2013vua}.

\begin{figure}[H]
  \centering
  \includegraphics[width=\textwidth]{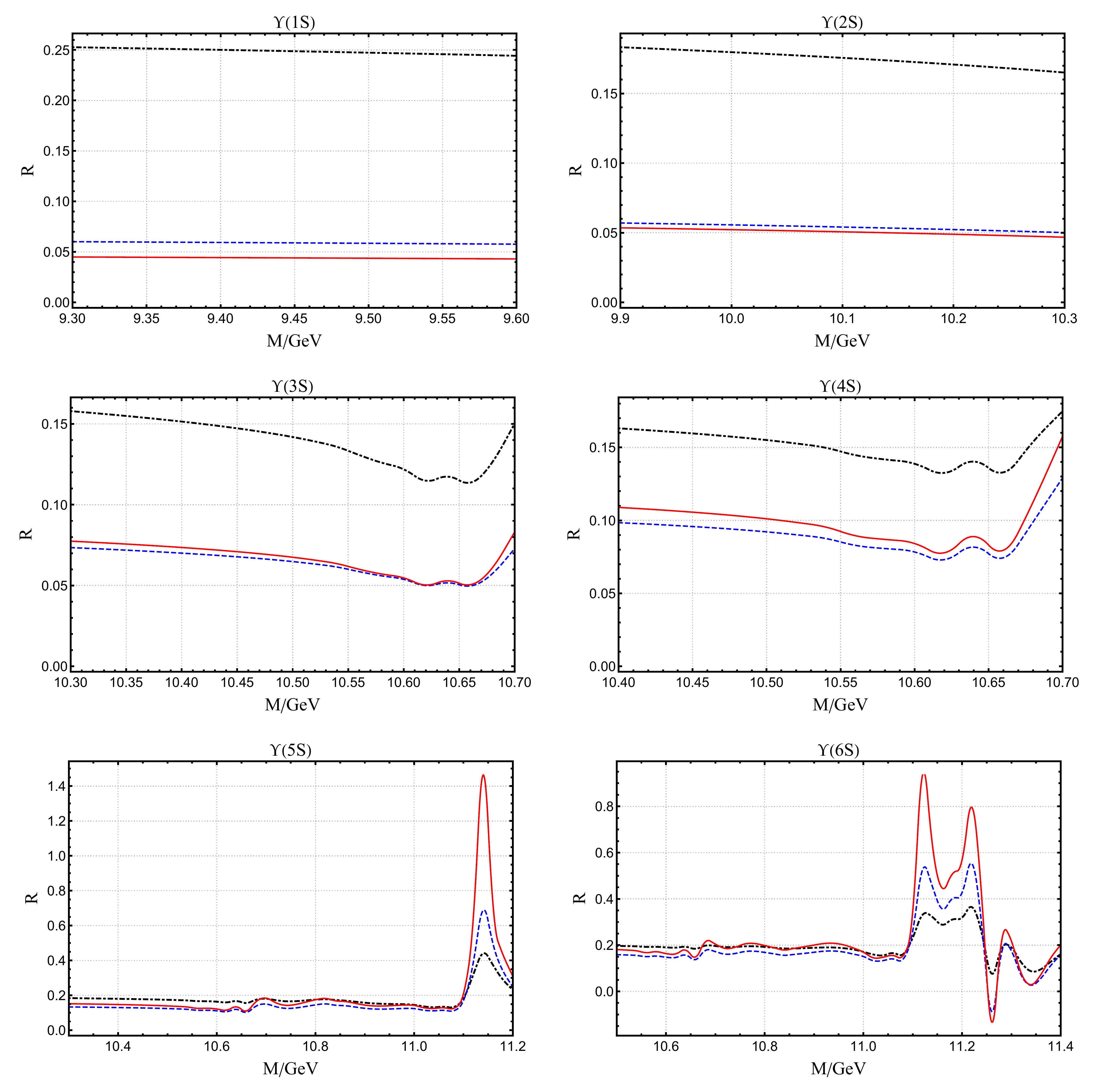}\\
  \caption{$R$ [defined in Eq.~\ref{Rdefine}] of the $\Upsilon$ family.
  Results corresponding to $r=0$ (no form factors),
  $r=0.335$fm (taken from~\cite{Ferretti:2013vua}), and $r=0.408$fm (our best fit) are denoted
  by black dot-dashed, blue dashed, and red solid curves, respectively.
  }
 \label{2SRatioCompare}
\end{figure}

From Fig.~\ref{2SRatioCompare},
one can clearly observe that the form factor is crucial to suppress the $B(2S)$'s contributions.
Without it, $B(2S)$ can contribute an additional $25\%$ of the $\Delta M$.
This deviation is not negligible if one wants to make a precise fit of the spectrum.
However, with a form factor $r=0.408$ fm,
the $\Delta M$ of $B(2S)$ is suppressed to less than $5\%$.
The previous question is partly answered,
the form factor is indispensable if one needs to suppress the contributions of $B(2S)$.
However, with the increase of $n$ in $\Upsilon(nS)$ family,
the form factor works less and less efficiently to suppress $R$.

Another effect of the form factor is that it adds more peaks to the oscillation of $\Delta M$.
This result has a pictorial explanation.
Since the form factor effectively serves as a cutoff,
this term will generally suppress every channel's contribution.
The key point is that different channels are suppressed by a different magnitude.
With the increase of the nodes of the wave functions of bottomonia and $B$ mesons,
at some specific energy point,
$B(2S)$'s relative contributions may increase.

\begin{figure}[H]
  \centering
  \includegraphics[width=\textwidth]{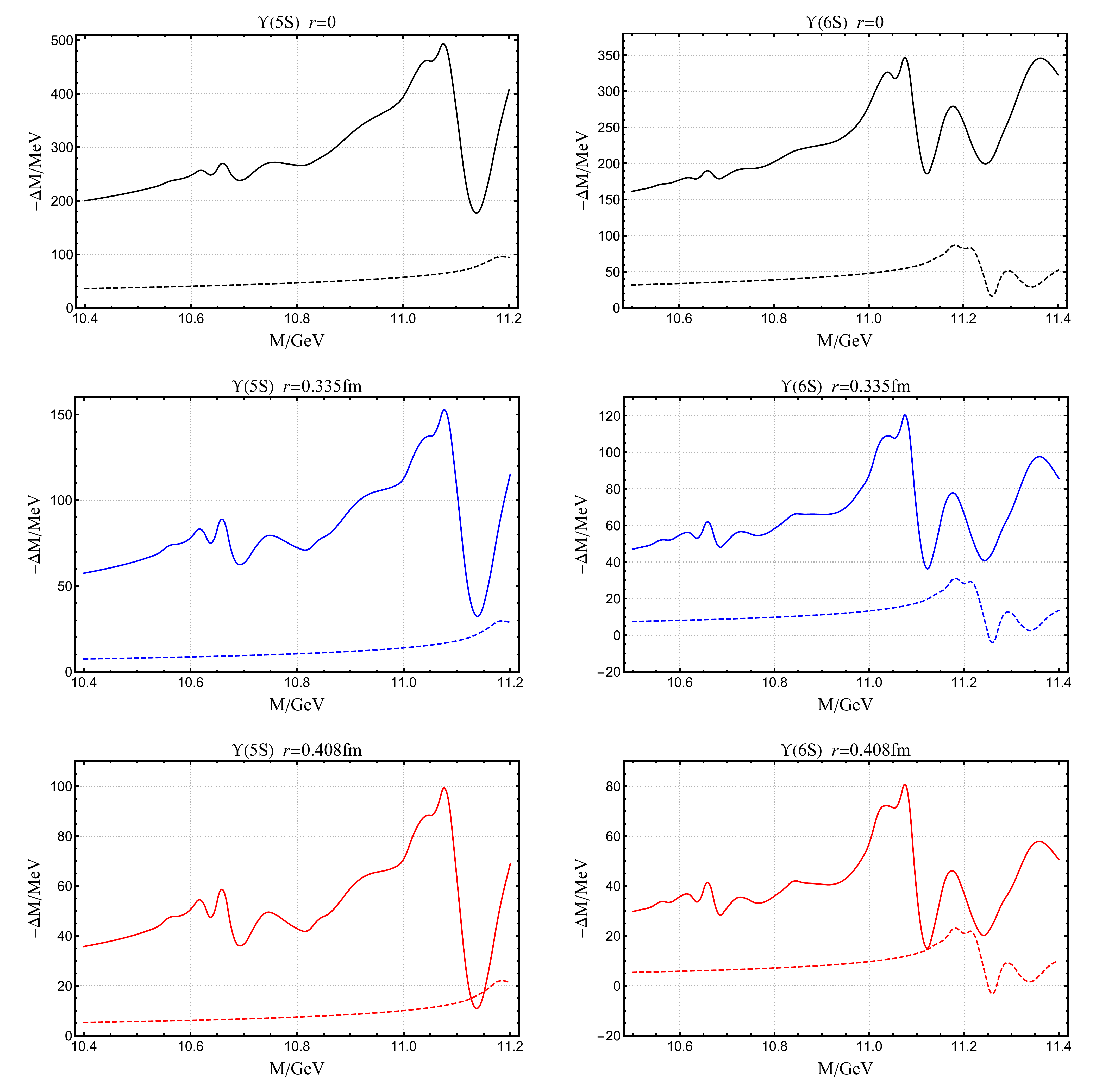}\\
  \caption{$-\Delta M$ of $\Upsilon(5S)$ and $\Upsilon(6S)$ with different form factors.
  Solid curves represent the denominator of Eq.~\ref{Rdefine},
  i.e., $\Delta M(1S,1S)+M(1S,1P)+M(1P,1P)$,
  and dashed curves represent the numerator, i.e. $\Delta M(2S,2S)+\Delta M(2S,1P)+\Delta M(2S,1S)$.
  $\Delta M$ corresponding to $r=0$, $r=0.335$fm, and $r=0.408$fm
  are denoted by black, blue, and red curves, respectively.
  The value of the $-\Delta M$ cannot compare within different form factors
  because the $\3P0$ coupling constants are not fit to reproduce experimental data.
  }
 \label{5S6SShift}
\end{figure}

Clearly the sharp peaks of $\Upsilon(5S)$ and $\Upsilon(6S)$ deserves a special analysis.
Both of them locate around $11.18$GeV.
Given the fact that the wave functions of $\Upsilon(5S)$ and $\Upsilon(6S)$ are quite different,
yet they both have the same sharp peaks at the same energy points,
it turns out to be more than just a coincidence.

$R$ is convenient to estimate $B(2S)$'s relative contributions;
however, it cannot tell whether the peak is due to the suppression of the denominator or
the increase of the numerator.
The complete information is encoded in $\Delta M$ itself.
In Fig.~\ref{5S6SShift},
we show $\Delta M$s for $\Upsilon(5S)$ and $\Upsilon(6S)$ with different form factors.

From Fig.~\ref{5S6SShift},
one can see that the enhancement of $R$ around $11.18$ GeV is in fact the bigger suppression
of $\Delta M(1S,1S)+\Delta M(1S,1P)+\Delta M(1P,1P)$, [or more precisely, the suppression of $\Delta M(1S,1P)$].
We have scanned $r$ in quite a wide range $0.17-1.08$ fm, which all show that
the form factor does not shift the positions of peaks or valleys of $\Delta M$.
Around this region, the form factor fails to suppress the $B(2S)$'s contribution is somewhat a coincidence,
because this suppression does not happen to $\Upsilon(5S)$ and $\Upsilon(6S)$ with Ref.\cite{Liu:2011yp}'s parameters;
their $\Delta M(1S,1P)$ always goes up.

Even in the low energy range where all the coupled-channels contribute a negative $\Delta M$,
no matter whether the form factor is considered or not,
the $B(2S)$ still contribute around $20\%$ of the total mass shift for $\Upsilon(5S)$ and $\Upsilon(6S)$.
In other words, the form factor fails to suppress $B(2S)$'s contributions.
This strongly indicates that only adding the Gaussian form factor
is not adequate to result in a complete calculation of the coupled-channel effects under the $\3P0$ framework.

One may argue that the coupled channel effect can be absorbed into the smeared potential
which returns wave functions broader than those of the Cornell model.
These broad wave functions indeed somewhat suppress the higher excited states contributions.
However, based on our experience, 
changing the size of the wave functions is also not able to solve the convergence issue.

We conclude this section with some comments on the $\3P0$ model.
In the classical $\3P0$ model,
the suppression due to the natural cutoff from wave functions
and the increase of the denominator of Eq.~\ref{mShift} are too weak.
Even with the remedy of the Gaussian form factor,
if one does not choose the cutoff $r$ carefully,
one will enhance instead of suppressing the $B(2S)$ contribution.
Compared with other models,
such as the microscopic decay model or the flux-tube breaking model,
which have rich microscopic details of the decay vertices,
the $\3P0$ model replaces these fine structures as an overall coupling constant $\gamma$.
This approximation may be not appropriate (see e.g. Ref.~\cite{Badalian:2012bt}).
In our case,
we show that this approximation leads to the bad convergence of the sum of the excited meson loops
for highly excited bottomonia.

The rich structure of the quark pair generation vertices may help to solve the convergence issue.
For example,
the Hamiltonian of the flux-tube breaking model suppresses the generation of the farther quarks,
one will expect that it is harder to generate excited intermediate states.

\begin{figure}[H]
  \centering
  \includegraphics[width=0.6\textwidth]{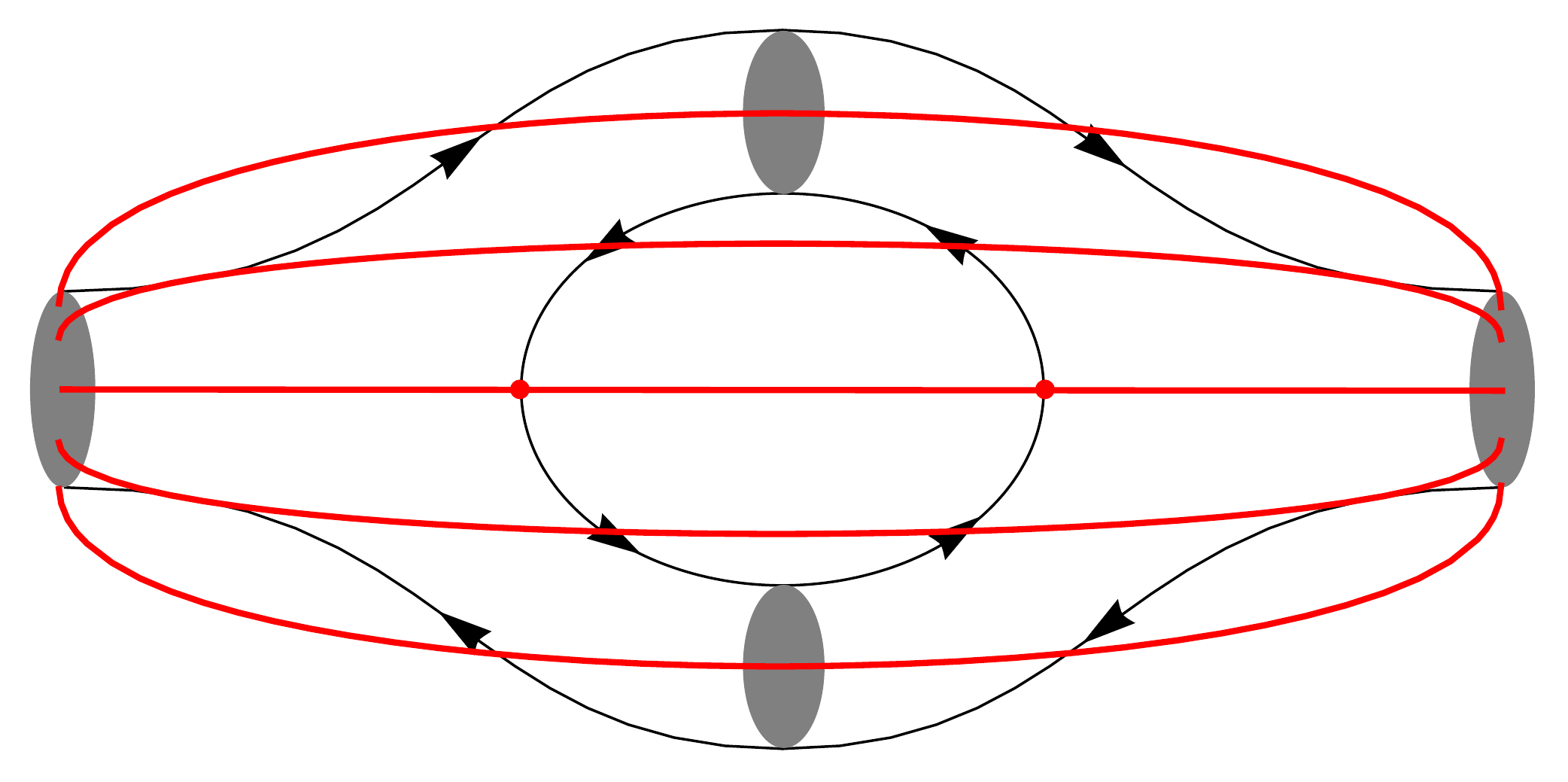}\\
  \caption{Sketch of coupled-channel effects in the flux-tube breaking model.
  Red lines stand for the flux-tube.
  The regions outside the flux-tube are shaved off,
  i.e., excited states' contributions in the loops are suppressed.
  }
 \label{CCofFTB}
\end{figure}

As illustrated in Fig.~\ref{CCofFTB},
contributions from the regions that are farther from the flux-tube line can be dropped,
as a result, the excited mesons' contribution is naturally suppressed.
Without this dynamical suppression,
it is more difficult to suppress $B(2S)$'s contributions even with the modified version of the $\3P0$ model,
where the form factor is added.

\section{Summary and Outlook}\label{summary}
In this paper, under the $\3P0$ model framework,
we explicitly calculated the excited $B$ mesons' contributions to the coupled-channel effects for the bottomonium.
We reveal the fact that compared to the ground state $B$ mesons,
contributions from $B(1P)$ mesons are generally the largest.
Up to this partial wave,
it is necessary to do the \emph{ab initio} calculations of the coupled-channel effects.

When we push the calculation beyond $B(1P)$,
we find some fundamental difficulties of the $\3P0$ model.
Even with the carefully chosen form factor,
it still cannot efficiently suppress intermediate state contributions of higher partial waves.
Since we do not fit our parameters with experimental data,
and we have exploited several different sets of parameters,
we have enough reasons to believe that the difficulties
are independent of the wave functions or the potential models.

We suggest that an efficient suppression mechanism such as a dynamically suppression
is needed to evaluate the coupled-channel effects.
How to effectively sum up all the intermediate loops of coupled-channels still remains to be an open issue.

\section*{Acknowledgements}
Yu Lu is grateful to Professor ~Feng-Kun~Guo for various discussions and suggestions
and to Mr.~Bao-Dong~Sun and Professor~Yu-Bing~Dong for lending their cluster computing resources.
M. Naeem~Anwar acknowledges several discussions with Professor~Cristoph~Hanhart during the Hadron Physics Summer School (HPSS 2016) at Forschungszentrum J\"ulich.
This work is supported by the National Natural Science Foundation of China under Grants No.~11261130311
(CRC110 by DFG and NSFC) and No.~11647601.
M. Naeem~Anwar is supported by CAS-TWAS President's Fellowship for International Ph.D. Students.

\providecommand{\href}[2]{#2}\begingroup\raggedright\endgroup

\end{document}